\documentclass[%
preprint,
superscriptaddress,
citeautoscript,
 amsmath,amssymb,
 aps,
prb,
floatfix,
]{revtex4-1}
\usepackage[version=3]{mhchem}
\usepackage{graphicx}
\usepackage{dcolumn}
\usepackage{bm}
\usepackage{multirow}
\usepackage{booktabs}
\usepackage{multirow}
\usepackage{xcolor}

\usepackage{pdfpages}
\usepackage{etoolbox} 

\makeatletter
\patchcmd{\@outputpage@head}{\@ifx{\LS@rot\@undefined}{}{\LS@rot}}{}{}{}
\makeatother

\begin{document}

\title{Enhanced screening and spectral diversity in many-body elastic scattering of excitons in two-dimensional hybrid metal-halide perovskites}

\author{F\'elix~Thouin}
\affiliation{School of Physics, Georgia Institute of Technology, 837 State Street NW, Atlanta, Georgia 30332, United~States}

\author{Daniele~Cortecchia}
\affiliation{Center for Nano Science and Technology@PoliMi, Istituto Italiano di Tecnologia, via Giovanni Pascoli 70/3, 20133 Milano, Italy}

\author{Annamaria~Petrozza}
\affiliation{Center for Nano Science and Technology@PoliMi, Istituto Italiano di Tecnologia, via Giovanni Pascoli 70/3, 20133 Milano, Italy}

\author{Ajay~Ram~Srimath~Kandada}%
\affiliation{Center for Nano Science and Technology@PoliMi, Istituto Italiano di Tecnologia, via Giovanni Pascoli 70/3, 20133 Milano, Italy}
\email[]{srinivasa.srimath@iit.it}

\author{Carlos~Silva}%
\affiliation{School of Chemistry and Biochemistry, Georgia Institute of Technology, 901 Atlantic Drive NW, Atlanta, Georgia 30332, United~States}
\affiliation{School of Physics, Georgia Institute of Technology, 837 State Street NW, Atlanta, Georgia 30332, United~States}
\email[]{carlos.silva@gatech.edu}


\date{\today}

\maketitle

\begin{bfseries}
	In two-dimensional hybrid organic-inorganic metal-halide perovskites, the intrinsic optical lineshape reflects multiple excitons with distinct binding energies~\cite{Gauthron2010,neutzner_exciton-polaron_2018},  each dressed differently by the hybrid lattice~\cite{thouin_phonon_2019}. Given this complexity, a fundamentally far-reaching issue is how Coulomb-mediated many-body interactions --- elastic scattering such as excitation-induced dephasing~\cite{moody_intrinsic_2015}, inelastic exciton bimolecular scattering~\cite{Xing2017}, and multi-exciton binding~\cite{Turner2010,thouin_stable_2018} --- depend upon the specific exciton-lattice coupling. 
	We report the intrinsic and density-dependent exciton pure dephasing rates 
	and their dependence on temperature by means of a coherent nonlinear spectroscopy. We find exceptionally strong screening effects on multi-exciton scattering relative to other two-dimensional single-atomic-layer semiconductors. Importantly, the exciton-density dependence of the dephasing rates is markedly different for distinct excitons. These findings establish the consequences of particular lattice dressing on exciton many-body quantum dynamics, which critically define fundamental optical properties that underpin photonics and quantum optoelectronics in relevant exciton density regimes.
\end{bfseries}

\begin{figure}[ht]
	\centering
	\includegraphics[width =\textwidth]{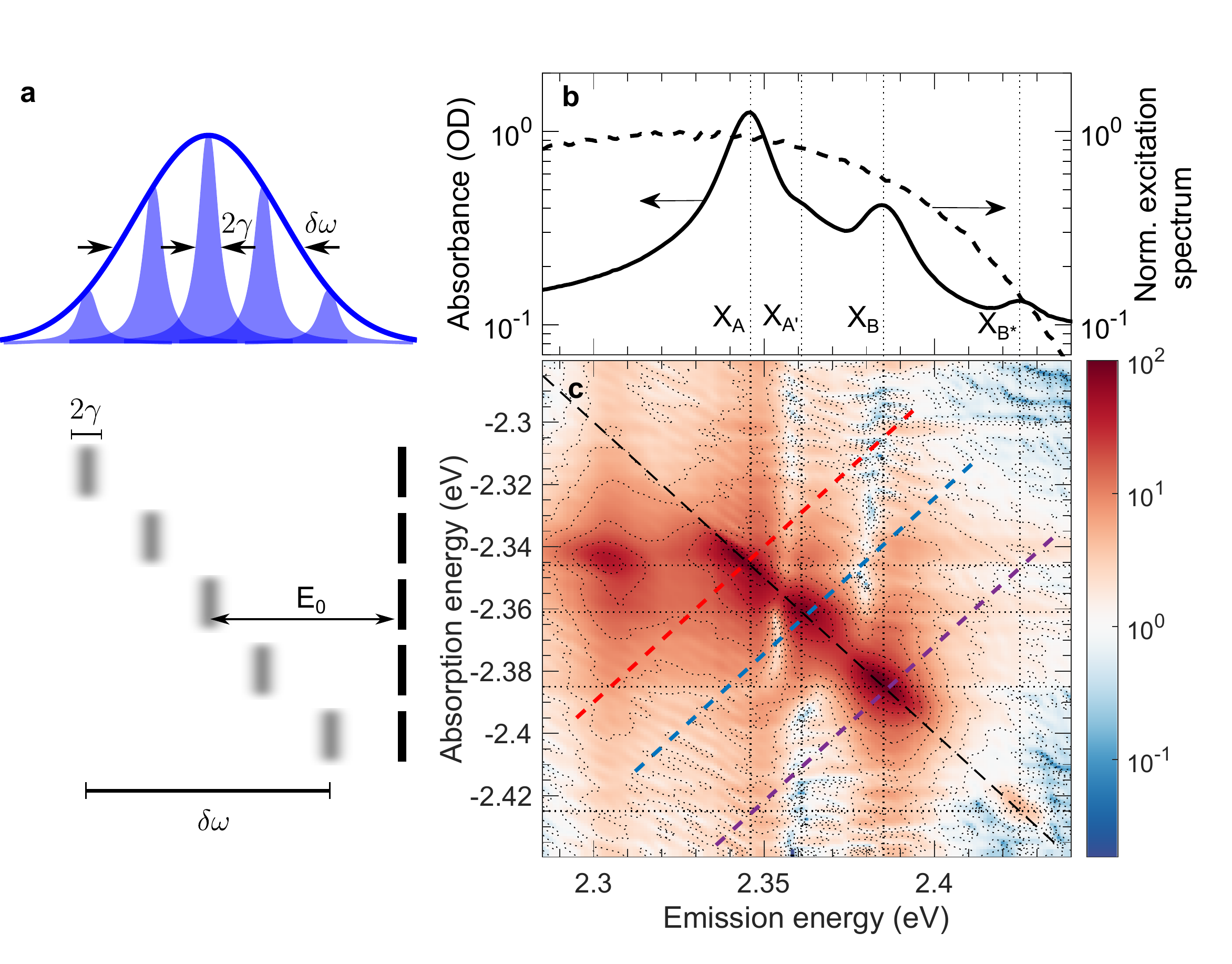}
	\caption{\textbf{Linear and two-dimensional coherent spectroscopy of \ce{(PEA)2PbI4} at 5\,K.} (a) Illustration of the inhomogeneous and homogeneous nature of line broadening. The total linewidth (blue line) is composed of a distribution of homogeneously broadened lines (blue areas). This arises from the simultaneous existence of uncorrelated transitions from the ground state (black lines) to short-lived excited states (blurred lines).  E$_{0}$ represents the central energy, $\delta\omega$ and $2\gamma$ the inhomogeneous and homogeneous linewidths respectively. (b) Absorption spectrum of \ce{(PEA)2PbI4} measured at 5\,K (black line) and normalized spectrum of the pulses used in 2D coherent excitation spectroscopy measurements (dashed line). Both scales are logarithmic. Dotted lines indicate the energy of excitons A, A$^{\prime}$, B, and B*, respectively, with increasing energy.(c) Absolute value of the 2D coherent rephasing spectrum of \ce{(PEA)2PbI4} measured at 5\,K with a pulse fluence of 40\,nJ/cm$^2$ and a pump-probe delay of 20\,fs. The color scale is logarithmic. Dotted lines indicate the energies of the aforementioned features. The paths of the diagonal cut (black dashed line) and anti-diagonal cuts at the diagonal energy of excitons A, A$^{\prime}$ and B (red, blue and purple dashed lines, respectively) are also shown. }  
	\label{fig1}
\end{figure}

Spectral transition linewidths provide pertinent insights into the system-bath interactions in materials because they depend on optical dephasing dynamics --- the processes by which the coherence that the driving electromagnetic wave imparts on the optical response dissipates due to scattering processes with lattice phonons, other excitations, and defects. Dephasing rates thus are very sensitive probes of the consequences of lattice dressing effects on excitons. Nevertheless, these are challenging to extract directly from linear optical probes such as absorption or photoluminescence spectroscopy given that the experimental linewidths typically arise from two distinct but co-existing contributions: homogenous and inhomogenous broadening (see Fig.~\ref{fig1}a). While the former is due to dephasing and is governed by the intrinsic finite lifetime of excited states and by dynamic disorder, the latter is caused by a statistical distribution of the transition energy due to static disorder, defects or grain boundaries. The exciton homogeneous linewidth 2$\gamma$ (full width at half maximum) is limited by the exciton lifetime ($\Gamma^{-1}$) and the dephasing rate mediated by exciton-exciton elastic scatterting (excitation-induced dephasing, $\gamma_{\mathrm{EID}}/\hbar$) and phonon scattering ($\gamma_{\mathrm{ph}}/\hbar$): 
\begin{equation}
\gamma=\frac{\hbar\Gamma}{2}+\gamma_{\mathrm{EID}}+\gamma_{\mathrm{ph}}.
\end{equation}
An accurate estimate of $\gamma$ is thus crucial to quantify the magnitude of the inter-exciton and exciton-phonon scattering cross-sections, which, in turn, influence the radiative rates and transport characteristics. 
Exciton-density-dependent transient photobleach linewidth analysis has been applied to address time-dependent lineshapes in a 2D hybrid perovskite~\cite{Abdel-Baki2016}. However, it is challenging to separate homogeneous dephasing effects from other $\vec{k}$-space-filling density effects by this means, and therefore to rigorously quantify $\gamma$. Here, we implement a nonlinear coherent spectroscopy to unambiguously extract the homogeneous dephasing rates of the various excitonic transitions in a polycrystalline thin film of single-layered \ce{(PEA)2PbI4} (PEA = phenylethylammonium). Fig.~\ref{fig1}b shows the linear absorption spectrum measured at 5\,K. Four distinct excitons (labelled A, A$^{\prime}$, B and B*) are observed about 200\,meV below the continuum band edge (see Supporting Information Fig.~S1 for the temperature-dependent absorption spectra). We quantify $\gamma$ as a function of exciton density and temperature and find that these different excitons display peculiar many-body phenomena.

Four-wave-mixing spectroscopies measure coherent emission due to a third-order polarization induced in matter by a sequence of phase-locked femtosecond pulses, and its dissipation reports directly on dephasing processes. In a two-dimensional coherent excitation geometry, one spectrally resolves this signal, and a 2D spectral correlation map of excitation/emission energies is constructed by time resolving the four-wave-mixing spectrum along a coherent excitation time variable and by Fourier transformation of the resulting coherence decay function. The absorption/emission diagonal axis thus contains the individual exciton resonances identified in the linear absorption spectrum, with any off-diagonal cross peaks displaying spectral correlations between them. With so-called rephasing phase matching, this 2D lineshape  permits separation of the homogeneous and inhomogeneous contributions to the optical linewidth. Specifically, if the homogeneous and inhomogeneous linewidths are comparable, which is the case for \ce{(PEA)2PbI4}~\cite{thouin_stable_2018,neutzner_exciton-polaron_2018}, the anti-diagonal lineshape is~\cite{siemens_resonance_2010,bristow_separating_2011}
\begin{equation}
S_{AD}(\omega_{ad})=\left|\frac{\exp\left(\frac{(\gamma-i\omega_{ad})^2}{2\delta \omega^2}\right)\text{erfc}\left(\frac{(\gamma-i\omega_{ad})}{\sqrt{2}\delta \omega}\right)}{\delta \omega (\gamma-i\omega_{ad})}\right|,
\label{antidiag}
\end{equation}
while the diagonal one reads as 
\begin{align}
&S_D(\omega_{d})= \sum_j \alpha_j \Bigg\vert \frac{\exp\left(\frac{(\gamma-i(\omega_{d}-\omega_j))^2}{2\delta \omega^2}\right)}{\gamma \delta \omega} \nonumber \\
\times& \left[\text{erfc}\left(\frac{(\gamma-i(\omega_{d}-\omega_j))}{\sqrt{2}\delta \omega}\right) 
+ \exp\left(\frac{2\gamma i (\omega_{d}-\omega_j)}{\delta \omega^2}\right) \text{erfc}\left(\frac{(\gamma+i(\omega_{d}-\omega_j))}{\sqrt{2}\delta \omega}\right)\right]
\Bigg\vert.
\label{diag_single}
\end{align}
Here $\text{erfc}$ corresponds to the complementary error function, $\omega_{ad}$ and $\omega_{d}$ are the anti-diagonal and diagonal angular frequencies, respectively, and $\delta \omega$ characterizes the inhomogeneous distribution. The sum in equation~\ref{diag_single} runs over the relative amplitudes $\alpha_j$ and the central diagonal energies $\omega_j$ of optical transitions of excitons A, A$^{\prime}$, B and B*. 

The laser spectrum used in this experiment, which covers all excitonic absorption features, is displayed 
in Fig.~\ref{fig1}b. 
The zero-population-time 2D map of the absolute value of the rephasing signal from (PEA)$_2$PbI$_4$ at 5\,K and a fluence of 40\,nJ/cm$^2$ (excitation density 8.8$\times 10^{18}$\,cm$^{-3}$ or 3.5$\times 10^{13}$\,cm$^{-2}$) is shown in Fig.~\ref{fig1}c. A schematic of our pulse sequence for this experiment is displayed in Fig.~S2, Supplemental Information, and the corresponding real components for different fluences are shown in Fig.~S3, as are details on the excitation density calculation. Four features are observed elongated along the diagonal axis, corresponding to excitons A, A$^{\prime}$, B and B*. 
Faint cross-peaks between all excitonic features are also observed, indicating that all the  transitions share a common ground state. An intense cross peak at an emission energy of 2.305\,eV is observed for exciton A despite the presence of a much weaker diagonal feature at lower energy than that dominant peak. Given the opposite phase of this feature (the real part of the spectra shown in Supplemental Information Fig.~S3) when compared to the corresponding diagonal peak, we attribute it to an excited-state absorption from a singly bound A exciton to a bound AA biexciton~\cite{thouin_stable_2018}. 

\begin{figure}[ht]
	\centering
	\includegraphics[width =0.5\textwidth]{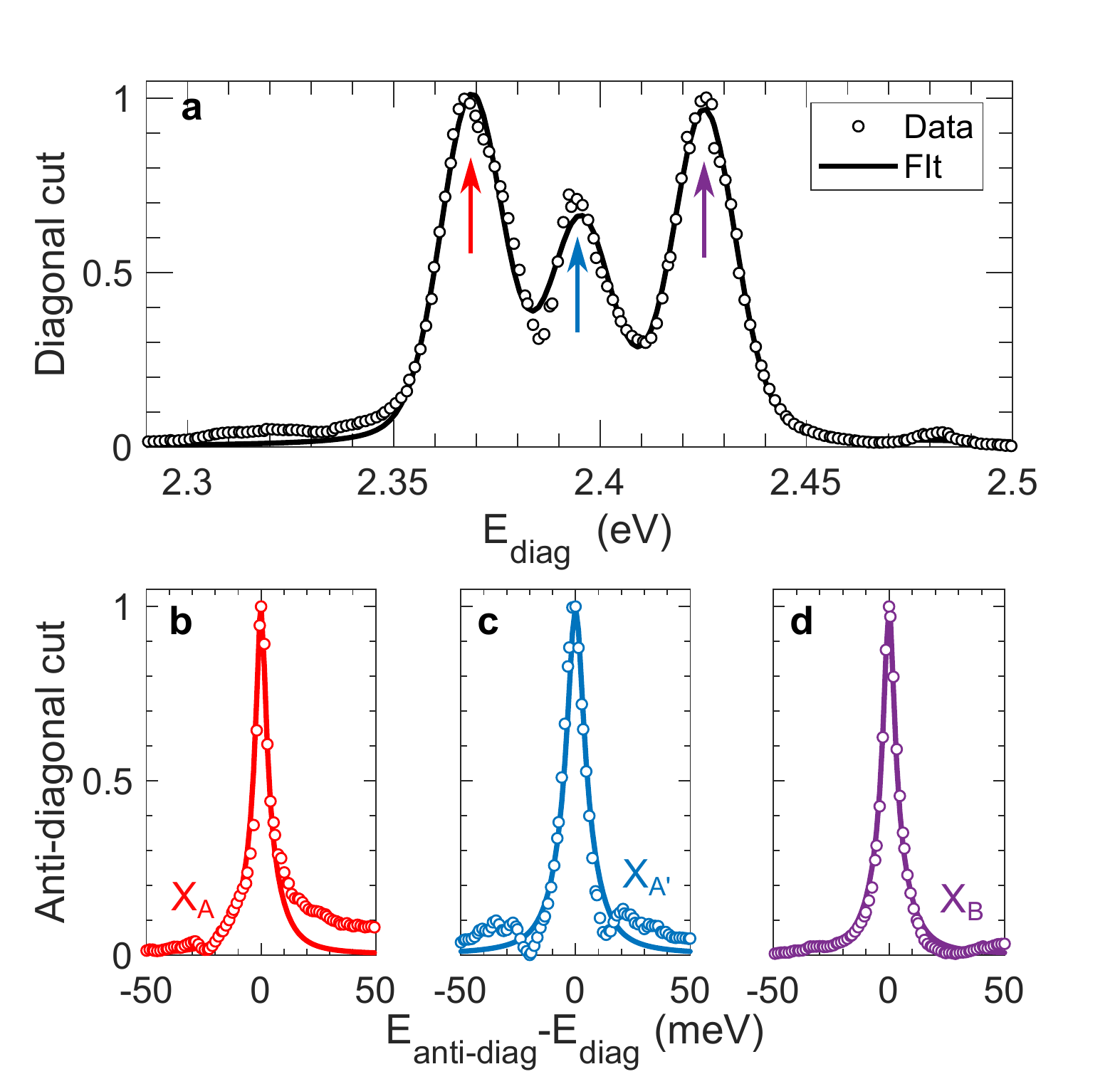}
	\caption{\textbf{Fitting diagonal and anti-diagonal spectral cuts to a lineshape model.} Normalized diagonal cut(a) and anti-diagonal cuts at diagonal energies of excitons A, A$^{\prime}$ and B (b, c and d respectively) of the two-dimensional coherent excitation rephasing spectrum presented in Fig.~\ref{fig1} c) (circles) and the result of the fitting procedure described in the main text (lines). Arrows in (a) mark the position along the diagonal where the antidiagonal cuts cross it. Due to the presence of low-amplitude cross peaks in the 2D spectrum, these also appear in the tails of the anti-diagonal cuts. To minimize their effect on the quality of the fits and to avoid overestimation of the linewidths, only points higher than 15\% of an anti-diagonal cut maxima were included in the fit. The quality of the fits presented here are representative of all the presented dataset. To test the robustness of the fits and to estimate the uncertainties on the extracted linewidths, we repeated the fitting procedure numerous times while adding white noise (5\% of the cut's maximum peak to peak) to the data.}
	\label{fig2}
\end{figure}

The diagonal and anti-diagonal cuts at the energies of excitons A, A$^{\prime}$ and B, along with the best fits to equations~\ref{diag_single} and \ref{antidiag}, are plotted in Fig.~\ref{fig2}a, b and c respectively. The only fit parameters are the amplitudes $\alpha_j$, homogeneous dephasing width $\gamma$ and inhomogeneous width $\delta \omega$ of each  optical transition. 

\begin{figure}[ht]
	\centering
	\includegraphics[width =\textwidth]{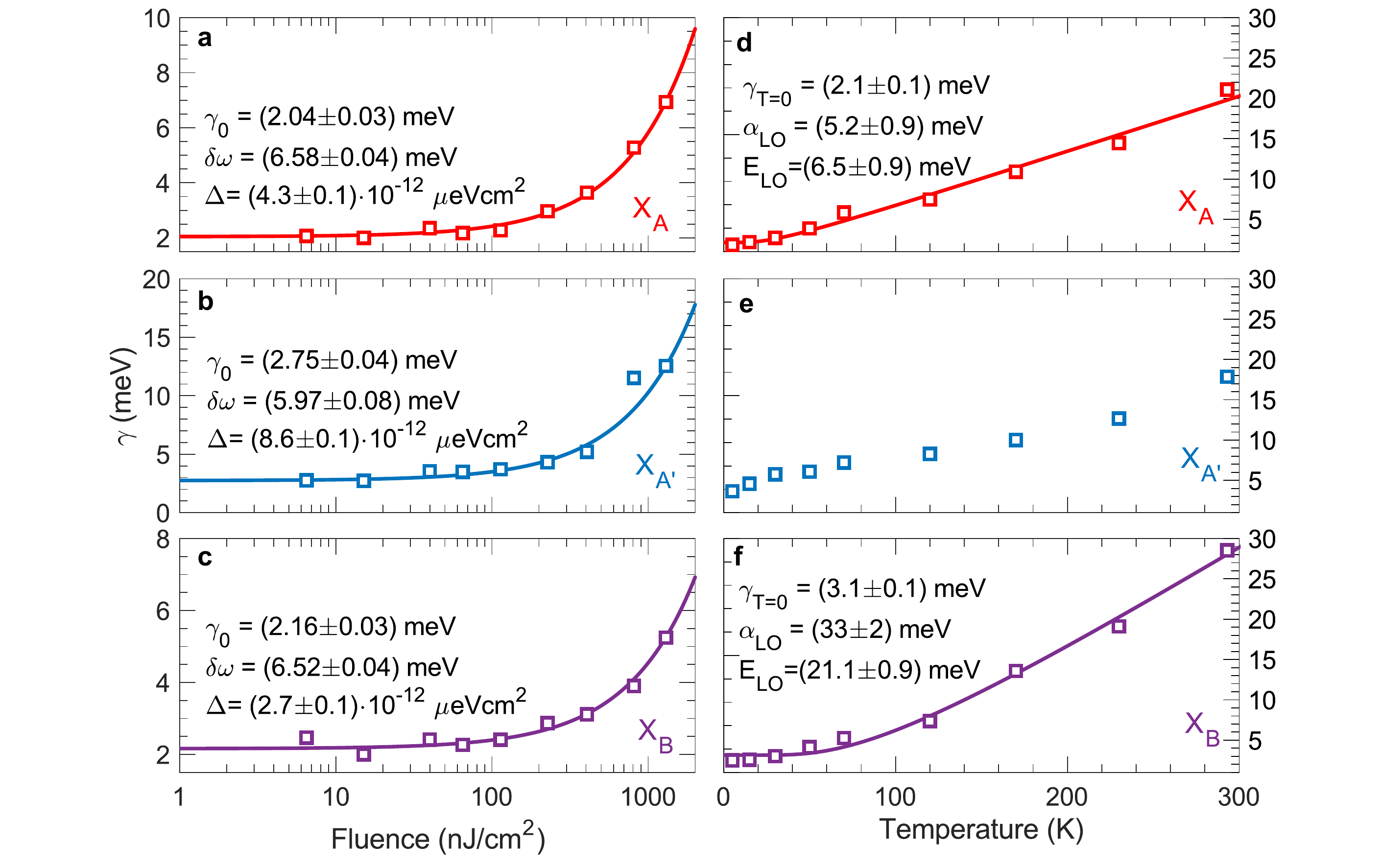}
	\caption{\textbf{Fluence and temperature dependence of the exciton dephasing rates.} Dephasing parameters $\gamma$ of excitons A, A$^{\prime}$ and B (a and d, b and e and c and f respectively) obtained from the simultaneous fitting of diagonal and anti-diagonal cuts, plotted as a function of excitation fluence (a, b and c) or temperature (d, e and f). Squares represent the experimental linewidths and lines are the best fit to the relevant model described in the main text. Error bars on the data are contained within the markers. For a, b and c, the sample's temperature is maintained at 5\,K while the excitation fluence is kept at 50\,nJ/cm$^2$ for measurements presented in panels d, e and f.}
	\label{fig3}
\end{figure}

To assess the contribution of many-body interactions on the dephasing dynamics of the different excitons, we acquired 2D coherent excitation spectra for a wide range of excitation fluences and sample temperatures (see Fig.~S4 and S5 of Supplemental Information for the raw data used here). 
The monotonic rise of $\gamma$ with the excitation fluence at 5\,K is shown in Fig.~\ref{fig3}. Such a dependence on exciton density $n$ is a consequence of broadening induced by exciton-exciton elastic scattering mediated by long-range Coulomb interactions:
\begin{equation}
\gamma_{\mathrm{EID}}(n)=\gamma_0+\Delta \cdot n.
\label{lin_gamma_powerdep}
\end{equation}
Here, $\gamma_0$ is the density-independent (intrinsic) dephasing rate and $\Delta$ is the exciton-exciton interaction parameter. Excitons in 2D metal-halide perovskites are confined to one of the inorganic quantum wells and are electronically isolated from the others due to the large inter-layer distance~\cite{thouin_phonon_2019} ($\sim 8$\,\AA~) imposed between them by the long organic cations. However, the sample itself, 40-nm thick, is composed of tens of these quantum wells, leading to a highly anisotropic exciton-exciton interaction. To allow for some form of comparison with other 2D semiconductors and quasi-two-dimensional quantum wells of similar thicknesses, we report the exciton-exciton interaction parameter, $\Delta$ in units of energy per area. The associated fits and the fit parameters are displayed in Fig.~\ref{fig3}a, b and c. While $\gamma_0$ is approximately 2\,meV with modest variation across the three probed excitonic transitions, $\Delta$ varies more substantially. It is $2.7\times 10^{-12}$\,$\mu$eV\,cm$^2$ for exciton B, increases to $4.3\times 10^{-12}$\,$\mu$eV\,cm$^2$ for exciton A,  and to $8.6\times 10^{-12}$\,$\mu$eV\,cm$^2$ for exciton A$^\prime$.  
It is not straight-forward to compare these values with those of other materials due to the ambiguities over the relevant values of the permittivity function and thus the Bohr radii. However, given the two-dimensional nature of the exciton and comparable exciton binding energies, monolayers of transition metal-dichalchogenides (TMDCs) provide a realistic benchmark. Intriguingly, previous 2D coherent excitation measurements on unencapsulated WSe$_2$~\cite{moody_intrinsic_2015} and encapsulated MoSe$_2$ ~\cite{martin2018encapsulation} revealed $\Delta = 2.7 \times 10^{-12} $\,meV\,cm$^2$ and $4 \times 10^{-13} $\,meV\,cm$^2$, respectively, three and two orders of magnitude higher than the value obtained here for \ce{(PEA)2PbI4}. This and the linearity of the dephasing rates over a wide range of excitation densities~\cite{ciuti_role_1998} highlights the substantial screening of the exciton-exciton interactions in these 2D perovskites. This is especially surprising given the high biexciton binding energy~\cite{thouin_stable_2018}, another characteristic that (PEA)$_2$PbI$_4$ shares with TMDC monolayers~\cite{You2015,kylanpaa_binding_2015}.   We will return to these differences in $\Delta$ below.

\begin{table}
	\begin{tabular}{cccc}
		\hline
		Polarization & $\gamma_{A}$ (meV) & $\gamma_{A^{\prime}}$ (meV) & $\gamma_{B}$ (meV)
		\\
		\hline
		Horizontal & 2.37$\pm$0.07 & 3.89$\pm$0.06 & 2.89$\pm$0.07\\
		Left circular & 2.70$\pm$0.10 & 4.35$\pm$0.08 & 3.27$\pm$0.07\\
		Right circular & 2.60$\pm$0.10 & 4.69$\pm$0.09 & 3.40$\pm$0.07\\
		\hline 
	\end{tabular}
	\caption{ Extracted $\gamma$ for excitons A, A$^{\prime}$ and B, using linearly or circularly polarized pulses at an excitation fluence of 50\,nJ/cm$^2$ and at 5\,K.}
	\label{Table}
\end{table}

To further highlight differences in multi-exciton elastic scattering behavior, we explore the dependence of the laser polarization state on $\gamma$, 
reported in Table~\ref{Table} at a fluence of 50\,nJ/cm$^2$ and 5\,K. For all excitons, $\gamma$ is consistently smaller with linearly polarized excitation than with circularly polarized pulses. Moreover, except for exciton A$^{\prime}$, it is independent on the helicity of the exciting pulses within our experimental uncertainty. When exciting the sample with linearly polarized pulses, excitons can scatter on both left and right circularly polarized excitons. However, they can only scatter with excitons of the same polarization when excited with circularly polarized pulses due to conservation of angular momentum~\cite{ciuti_role_1998}. Table~\ref{Table} further points to differences in scattering behavior of distinct excitons.

We now turn to the temperature dependence of $\gamma$ to investigate exciton-phonon interactions. The measured $\gamma$ for excitons A, A$^{\prime}$ and B as a function of temperature are presented in Fig.~\ref{fig3}d, e and f, respectively. We fit this data with 
\begin{equation}
\gamma_{\mathrm{ph}}(T)=\gamma_{T=0}+\alpha_{\mathrm{LO}}\left[\frac{1}{\exp(E_{\mathrm{LO}}/k_BT)-1}\right],
\label{therm_broad}
\end{equation}
which assumes that line broadening arises from scattering of excitons with a single thermally populated phonon mode of effective energy $E_{\mathrm{LO}}$ with an effective interaction parameter $\alpha_{\mathrm{LO}}$. The parameters representing the best fit for each exciton are shown in Fig.~\ref{fig3}d and f. 
This model approximates well the behavior of excitons A and B but fails to reproduce that of exciton A$^{\prime}$, an indication of the presence of complex lattice interactions as elucidated in our recent works~\cite{thouin_stable_2018, neutzner_exciton-polaron_2018, thouin_phonon_2019}. To compare the strength of the obtained interaction parameters with those of other semiconductors, we obtain the linear interaction parameter $\Delta_{\mathrm{ph}}$ by taking the low-temperature limit of equation~\ref{therm_broad}, yielding $\Delta_{\mathrm{ph}}=\alpha_{\mathrm{ph}}k_B/E_{\mathrm{LO}}$. This gives linear interaction parameters of $70 \pm 20$ and $130\pm10$\,$\mu$eV/K for exciton A and B, respectively. These are almost an order of magnitude larger than those in two-dimensional quantum wells~\cite{honold_collision_1989,wagner_coherent_1997} and curiously comparable to monolayers of (covalent) TMDCs~\cite{moody_intrinsic_2015} given the ionic character of the lattice and the resulting strong electron-phonon coupling.

Despite the strong polaronic dressing of the excitons with distinct yet multiple phonon modes~\cite{thouin_phonon_2019}, here we find that only one phonon effective energy dominates the scattering process via weak interactions. Exciton A, dressed primarily by the in-plane lattice modes~\cite{thouin_phonon_2019}, interacts with optical phonons at 6.5\,meV, which correspond to lattice modes within the inorganic lead-halide network. Exciton B, which is predominantly dressed by out-of-plane lattice modes~\cite{thouin_phonon_2019} (see also ref.~\citenum{fieramosca_tunable_2018}), interacts with higher-energy phonons at 21\,meV, which we have identified to have large contributions from the motion of the organic cation, particularly from the $\pi$--$\pi$ motion of the phenyl groups~\cite{dragomir2018lattice}. This 
behavior only highlights the complex phonon coupling scenario which may be beyond a conventional Fr\"ohlich formalism and calls for further experimental and theoretical investigations.

We consider that polaronic effects rationalize the two peculiar observations of this work in the context of many-body interactions in 2D perovskites --- the remarkably low values of $\Delta$, and non-trivial differences of this parameter and $\Delta_{\mathrm{ph}}$ for diverse exciton resonances. Remarkably low inter-exciton scattering rates along with relatively weak exciton-phonon interactions may be attributed to a polaronic protection mechanism~\cite{zhu_charge_2015,miyata_large_2017}, where Coulomb interactions that are at the heart of both of these scattering events are effectively screened by the dynamic lattice motion. Such a mechanism has been invoked in the case of bulk 3D perovskites to explain slow cooling of hot carriers~\cite{joshi_dynamic_2019} and long carrier lifetimes~\cite{wehrenfennig_charge-carrier_2014, brenner_are_2015}, and analogous comparisons can be drawn in their 2D counterparts. Excitons reorganize the lattice along well-defined configurational coordinates~\cite{thouin_phonon_2019}, which  dresses the exciton with a phonon-cloud. Although we currently do not have a reliable estimate of the spatial extent of this deformation, i.e., the polaron radius, based on the polaron coupling constants in lead-halide perovskites~\cite{miyata_large_2017}, we may hypothesize that the radius is much larger than the exciton radius. Such polaronic contributions effectively screen exciton-exciton interactions. At higher densities, we expect a high probability of multiple excitons within the polaron radius, in which case biexcitons may be populated given their high binding energy~\cite{thouin_stable_2018}. 

Lastly, we note that the zero-density ($\gamma_0$) and the zero-temperature ($\gamma_{T=0}$) dephasing widths for all the excitonic transitions are between 2--3\,meV, which corresponds to a pure dephasing time $T_2^* = \hbar /\gamma_{0,T=0} \approx$ 500\,fs. Time-resolved photoluminescence measurements performed at 5\,K have revealed a lower limit for the exciton lifetime to be around 100\,ps~\cite{straus2016direct}, suggesting a radiative width $\gamma_{\mathrm{rad}} \sim 0.04$\,meV, which is much lower than the measured lower-limit $\gamma_{0,T=0}$. This clearly demonstrates the presence of an additional dephasing mechanism in addition to inter-exciton and phonon scattering, possibly due to the defective nature of the polycrystalline film, the presence of other degenerate dark states which excitons scatter in and out of, or via low-energy acoustic phonons unresolved in the current experiment. We note that similar subpicosecond dephasing times were measured at low temperature for photocarriers in \ce{CH3NH3PbI3}, which were noticed to be a factor of $\sim 3$ times longer than in bulk GaAs~\cite{March2017}. This was ascribed to weaker Coulomb interactions in that perovskite, limiting the role of excitation-induced dephasing effects compared to III-V semiconductors. In ref.~\citenum{March2017}, it was speculated that dynamic large polarons account for the relatively long photocarrier dephasing time. We point out that these relatively slow dephasing rates are still much too fast for quantum optoelectronics applications such as single-photon emitters~\cite{utzat2019coherent}, which require coherence times approaching the radiative lifetime.

In a general sense, multi-exciton interactions are determined by an interplay of Coulomb forces, and in ionic crystalline systems, the role of the lattice in mediating them is of fundamental importance. We have focused on the comparison of elastic scattering processes of spectrally distinct excitons within the excitation fine structure, manifested via the dephasing rate. In a previous publication, we have reported high biexciton binding energies in this material~\cite{thouin_stable_2018}, and importantly, that biexcitons display distinct spectral structure. We highlight that in ref.~\citenum{thouin_stable_2018}, the biexciton binding energy of excitons A and B appears to be different, and exciton A displays clear evidence of repulsive interactions in the two-quantum, two-dimensional spectral correlation map. Such interactions might give rise to inelastic scattering of exciton A, perhaps related to Auger recombination~\cite{Xing2017}. These overlapping dynamics would be deterministic in biexciton lasers~\cite{booker2018vertical}, for example, if these devices were to be rigorously implemented. The extent to which the spectral scattering rates depend on spectral structure might also determine the dynamics of exciton polaritons in semiconductor microcavities, in which quantum fluids are formed by polariton-polariton inelastic scattering~\cite{Carusotto2013}. It has been demonstrated that polariton-polariton interactions in 2D lead-halide perovskites are strong~\cite{fieramosca2018two}, such that the $\sim 0.5$-ps intrinsic dephasing times reported here are long compared to Rabi oscillation periods given $\gtrsim 150$-meV Rabi splittings~\cite{brehier_strong_2006}.  In the search for room-temperature polartion condensates, it is important to know how polaronic effects control exciton many-body dephasing dynamics, and how hybrid perovskites differ in this respect from other two-dimensional candidate semiconductor systems such as monolayer TMDCs~\cite{moody_intrinsic_2015,martin2018encapsulation}. If polaronic effects mitigate excitation induced dephasing in 2D hybrid perovskites, then the quantum dynamics in a vast range of structurally diverse derivatives merit profound experimental and theoretical investigation.
  
\section*{Methods}
\subsection*{Sample preparation}
Thin films of \ce{(PEA)2PbI4} (thickness of 40\,nm) were prepared on sapphire substrates (optical windows 25\,mm$\times$0.5\,mm, Crystran) by spin coating a 0.05\,M solution of the perovskite in N,N-Dimethylformamide (DMF). 12.5\,mg of \ce{(PEA)I} (Dyesol) was mixed with 11.5\,mg \ce{PbI2} (TCI) and dissolved in 500\,$\mu$l of DMF (Sigma Aldrich, anhydrous, 99.8$\%$). The solution was left to dissolve on a hotplate at 100$^{\circ}$C for 1 hour. After exposing the substrate under oxygen plasma, the solution (kept at 100$^{\circ}$C) was spin coated on the sapphire window at 6000 rpm, 30 s, and the film was annealed on a hotplate at 100$^{\circ}$C for 15 minutes. Solution and film preparation were performed in a glove-box under \ce{N2} inert atmosphere.

\subsection*{Multidimensional spectroscopy}
The pulses used for these measurements were generated by a home-built single pass non-collinear optical parametric amplifier pumped by the third harmonic of a Yb:KGW ultrafast laser system (Pharos Model PH1-20-0200-02-10, Light Conversion) emitting 1,030-nm pulses at 100\,kHz, with an output power of 20 W and pulse duration of 220\,fs. Multidimensional spectroscopic measurements were carried on using a home-built implementation of 
a pulse-shaper-based multidimensional spectrometer that passively stabilizes the relative phase of each pulses~\cite{Turner2010}. Our experimental implementation is described in detail in ref.~\citenum{thouin_stable_2018}, albeit with a different ultrafast laser source. Each beams was independently compressed using chirp-scan~\cite{VincentL2013} to a pulse duration of 25\,fs FWHM, measured using cross-correlated second harmonic frequency resolved optical gating (SH-XFROG) in a 10\,$\mu$m-thick BBO crystal placed at the sample position. A typical SH-XFROG trace is shown in Fig.~S6 of Supplemental Information. The sample was kept at cryogenic temperatures using a vibration-free cold-finger closed-cycle cryostat from Montana Instruments.

\begin{acknowledgements}
	A.R.S.K.\ acknowledges funding from EU Horizon 2020 via a Marie Sklodowska Curie Fellowship (Global) (Project No.\ 705874). This work is partially supported by the National Science Foundation (Award 1838276). C.S.\ acknowledges support from the School of Chemistry and Biochemistry and the College of Science of Georgia Institute of Technology.
\end{acknowledgements}

\section*{Author Contributions}
F.T. carried out all measurements and performed the analysis of the experimental data. D.C.\ synthesized the samples. A.P.\ supervised the sample preparation activity. C.S.\ and A.R.S.K.\ supervised the ultrafast spectroscopy activity. A.R.S.K.\ and C.S.\ conceived the project. All authors contributed to the redaction of the manuscript. C.S.\ and A.R.S.K.\ are to be considered corresponding co-authors.

\newpage
\clearpage


\section*{Supplementary material}

# Supplemental Information: Enhanced screening and spectral diversity in many-body elastic scattering of excitons in two-dimensional hybrid metal-halide perovskites

Félix Thouin,[1] Daniele Cortecchia,[2] Annamaria Petrozza,[2] Ajay Ram Srimath Kandada,[2] and Carlos Silva[3, 1]

[1]*School of Physics, Georgia Institute of Technology,*
*837 State Street NW, Atlanta, Georgia 30332, United States*
[2]*Center for Nano Science and Technology@PoliMi,*
*Istituto Italiano di Tecnologia, via Giovanni Pascoli 70/3, 20133 Milano, Italy*
[3]*School of Chemistry and Biochemistry,*
*Georgia Institute of Technology, 901 Atlantic Drive NW,*
*Atlanta, Georgia 30332, United Sates*

(Dated: May 11, 2019)

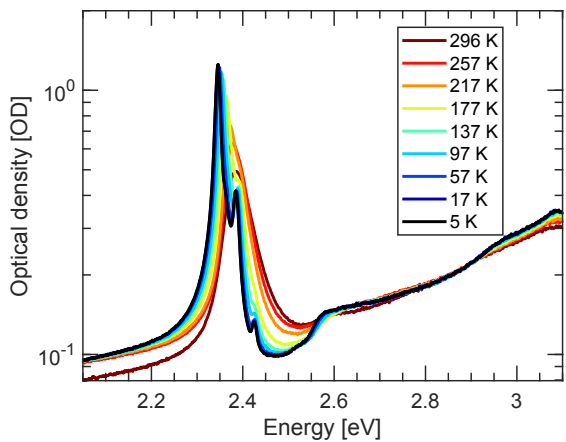

FIG. S1. Temperature dependent absorption spectra on a logarithmic scale.

## ESTIMATION OF THE INDUCED EXCITATION DENSITY

The presence of spectral features in the absorption spectra as well as the pulses' non-uniform spectra must be accounted for when calculating the excitation density induced by one pulse. We do so by calculating the fluence to surface excitation density factor $\Theta$ as

$$\Theta = \int_0^\infty \frac{A(E)P(E)}{E} dE \qquad (1)$$

where $A(E)$ is the absorption spectrum, $P(E)$ is the pulse spectrum normalized so that its integral yields unity and $E$ the energy. To obtain the surfacic excitation density, we then multiply the pulse fluence by $\Theta$. The volumic excitation density is then obtained by dividing this number by the thickness of the material, namely 40 nm. We obtained a value of $8.75 \times 10^{20}$ excitons per J for $\Theta$.

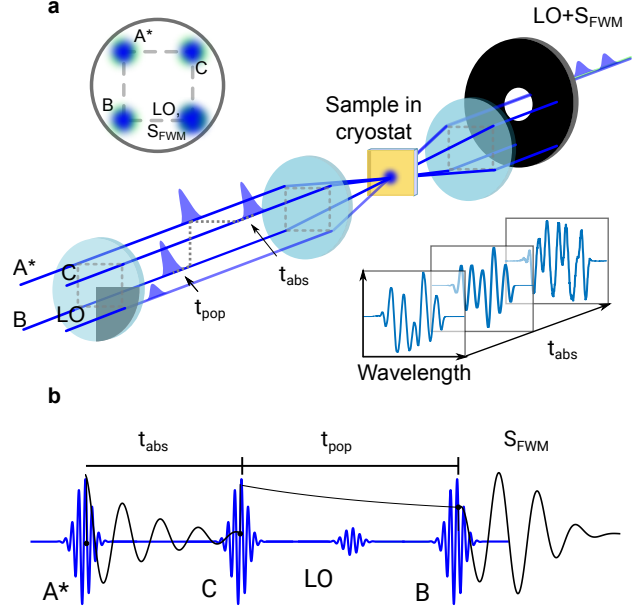

FIG. S2. (a) Schematic of rephasing multi-dimensional spectroscopy experiment. In the BOXCARS geometry (upper left corner), four copropagating beams are arranged in the corners of a square. One beam, the local oscillator (LO), is attenuated before reaching the sample so that it does not interact non-linearly with the sample. After the other pulses create a non-linear polarization in the sample, the latter radiates a four-wave mixing signal in the direction of the LO. Their spectral interference is then acquire by a spectrometer so that the signal's full electric field is resolved through heterodyne detection. The signal is heterodyne detected for each relevent time step. (b) Pulse sequence used for a rephasing measurement. The blue line represents the applied electric field by the pulse sequence and the black lines represent the value of the relevent element of the system's density matrix as a function of time (going from left to right). The beam opposed to the LO interacts first (A* in this case), creating a an oscillating coherence. The second beam, (C in thise case) later maps this coherence back to a population before the last beam (B in this case) forces the system into another coherence. This oscillation of a coherence between states connected by an electric dipole interaction radiates the detected signal.

3

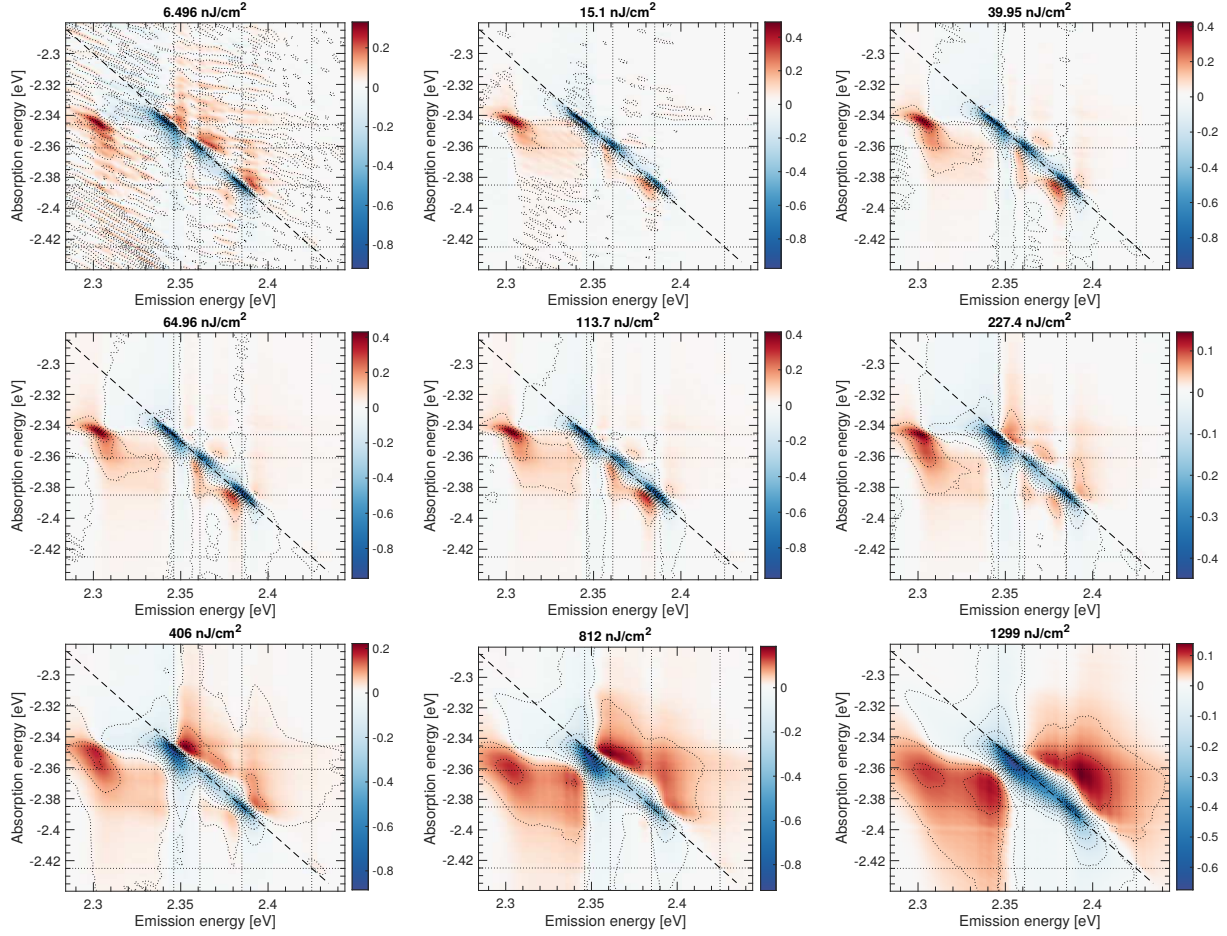

FIG. S3. Real part of fluence dependent photon-echo spectra at 5 K. The signal was phased at an emission energy of 2.385 eV. Each spectrum was normalized by the maximal value of its norm.

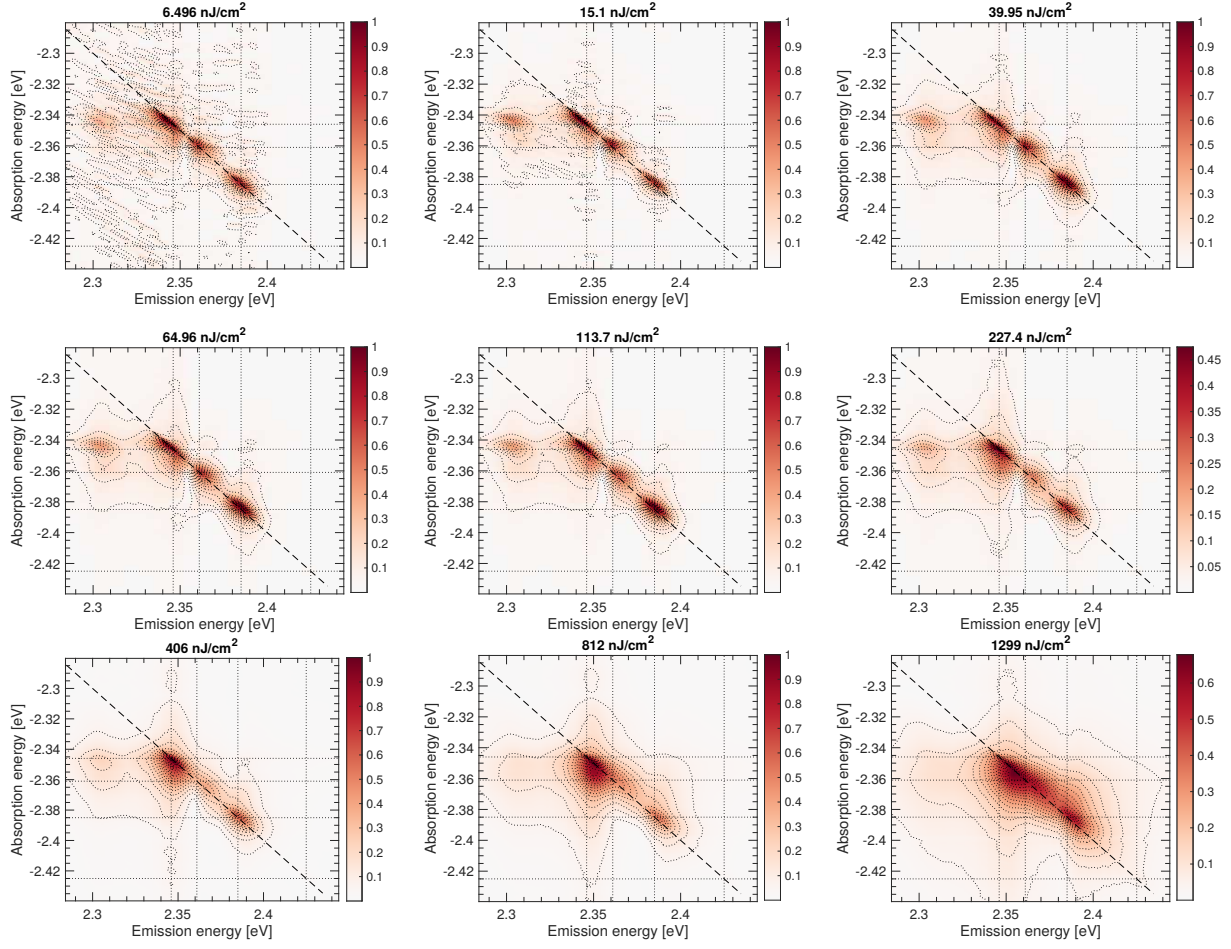

FIG. S4. Norm of fluence dependent photon-echo spectra at 5 K. Each spectrum was normalized by the maximal value of its norm.

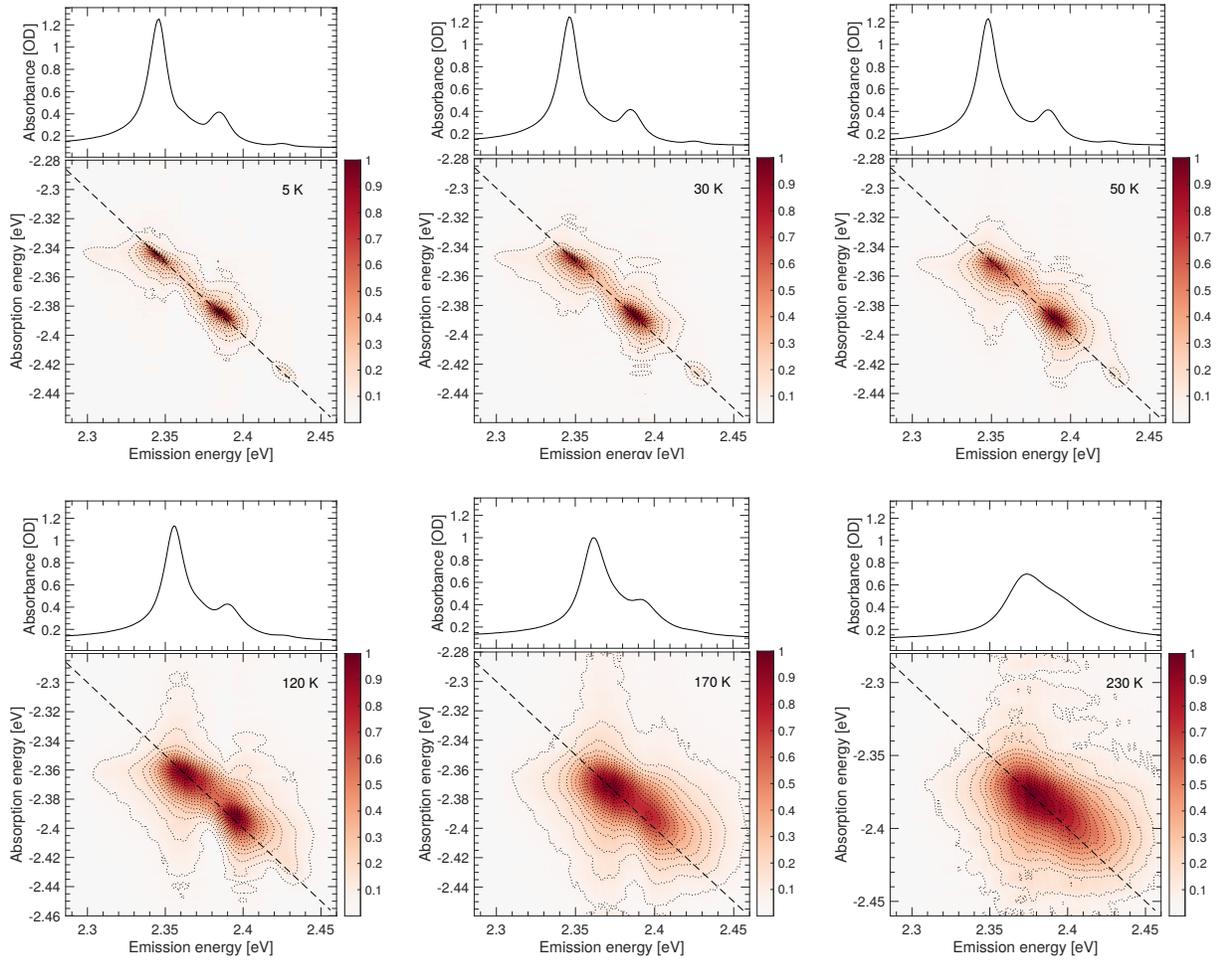

FIG. S5. Norm of temperature dependent photon-echo spectra at 5 K. Each spectrum was normalized by the maximal value of its norm. The corresponding absorption spectrum is plotted above each photon-echo spectra.

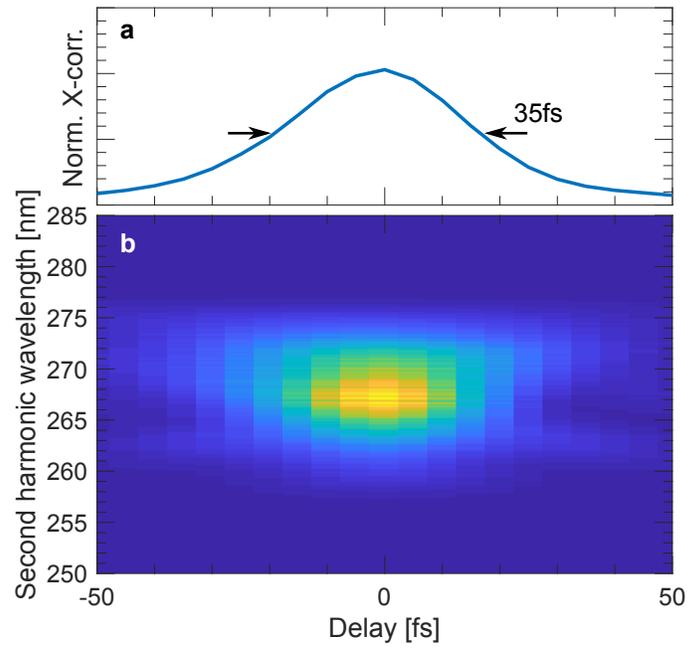

FIG. S6. (a) Cross-correlation of one the pulses used in the experiments with the local oscillator pulse. it is obtained by spectraly integrating the SH-XFROG measurement. The FWHM of the cross-correlation is 35 fs, yielding a pulse duration of 25 fs. (b) SH-XFROG trace of one of the pulses with the LO pulse.



\end{document}